\documentclass[journal=mamobx,manuscript=article]{achemso}

\usepackage{amsmath,amssymb,amsfonts}
\usepackage{bm} 
\usepackage{graphicx}
\usepackage{caption}
\usepackage{subcaption}
\usepackage{xcolor}
\usepackage{tikz}
\usepackage[normalem]{ulem}
\usepackage[hidelinks]{hyperref}
\graphicspath{{figures/}}

\newcommand{\jAA}{J_{AA}^{{II}}}
\newcommand{\jBB}{J_{BB}^{{II}}}
\newcommand{\jAB}{J_{AB}^{{II}}}
\newcommand{\jBA}{J_{BA}^{{II}}}

\newcommand{\jA}{J_{A_S}}
\newcommand{\jB}{J_{B}}
\newcommand{\hAL}{H_{A_L}}
\newcommand{\hA}{H_{A_S}}
\newcommand{\hB}{H_{B}}
\newcommand{\gA}{G_{A_S}}
\newcommand{\gB}{G_{B}}
\newcommand{\rhoqA}[2]{\rho_{#1}^{#2}}
\newcommand{\ralpha}[1]{\bm{#1}_{l}^{\alpha}}

\author{Merin Joseph}
\email{fsmj@leeds.ac.uk}
\author{Daniel J. Read}
\email{D.J.Read@leeds.ac.uk}
\author{Alastair M. Rucklidge}
\email{A.M.Rucklidge@leeds.ac.uk}
\affiliation[University of Leeds]
{School of Mathematics, University of Leeds, Leeds LS2 9JT, UK}

\title[Two length scale block copolymers]{Design of linear block copolymers and $ABC$ star terpolymers that produce two length scales at phase separation}

\keywords{Quasicrystals, Phase separation, block copolymer, Self assembly, Random Phase Approximation}

\begin{document}

\begin{tocentry}


\includegraphics[scale=1]{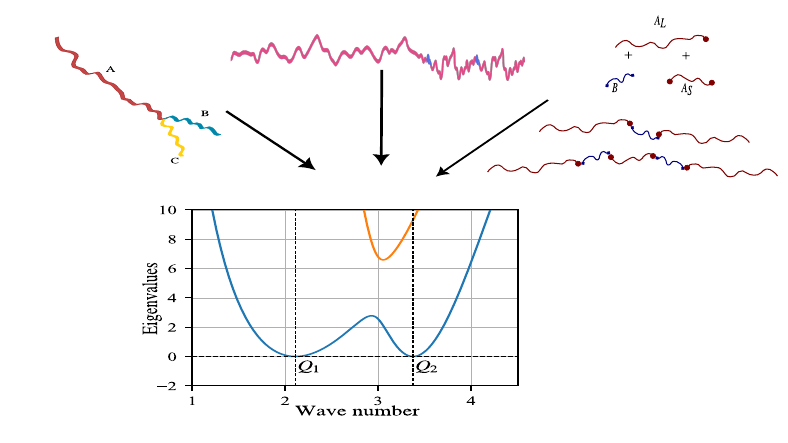}


\end{tocentry}

\begin{abstract}
 Quasicrystals (materials with long range order but without the usual spatial periodicity of crystals) were discovered in several soft matter systems in the last twenty years. 
 The stability of quasicrystals has been attributed to the presence of two prominent length scales in a specific ratio, which is 1.93 for the twelve-fold quasicrystals most commonly found in soft matter.
 We propose design criteria for block copolymers such that quasicrystal-friendly length scales emerge at the point of phase separation from a melt, basing our calculations on the Random Phase Approximation.
 We consider two block copolymer families: linear chains containing two different monomer types in blocks of different lengths, and $ABC$ star terpolymers.
 In all examples, we are able to identify parameter windows with the two length scales having a ratio of~1.93.
 The models that we consider that are simplest for polymer synthesis are, first, a monodisperse $A_{L}BA_{S}B$ melt and, second, a model based on random reactions from a mixture of $A_L$, $A_S$ and $B$ chains: both feature the length scale ratio of~1.93 and should be relatively easy to synthesise.
 \\
 \textit{Macromolecules} \textbf{2023}, \textit{56}, 7847--7859, \href{https://doi.org/10.1021/acs.macromol.3c00800}{10.1021/acs.macromol.3c00800} (Open Access).
\end{abstract}

\section{Introduction}

Quasicrystals are crystals that have long range order, and so have sharp X-ray diffraction spectra, and yet do not have the spatial periodicity usually associated with crystals\cite{Lifshitz2007a}.
They often have rotation symmetries that are incompatible with spatial periodicity. 
The first examples were in metal alloys and had icosahedral symmetry\cite{Shechtman1984a}.
Subsequently Zeng \textit{et al.}\cite{Zeng2004} reported quasicrystals in micelles made from wedge-shaped dendrimers: these examples were quasicrystalline, with twelve-fold rotation symmetry in two dimensions and periodic in the third.
Twelve-fold quasicrystals have recently been reported in systems as simple as oil--water--surfactant mixtures\cite{Jayaraman2021}.
An earlier discovery by Hayashida \textit{et al.}\cite{Hayashida2007} was an example with the same dodecagonal rotation symmetry in a three-component $ABC$~star terpolymer blend of polyisoprene, polystyrene and poly(2-vinylpyridine). 
In this case, the bulk properties were as if the three components were immiscible, forming tiles composed the $A$, $B$ and $C$ block copolymers in two dimensions, with the junction points aligned in the third.
The small-angle X-ray scattering pattern of the quasicrystal sample had two circles of wavevectors with twelve peaks on each circle, demonstrating dodecagonal rotation symmetry and the presence of two length scales, roughly in a ratio of $1:1.93$.

Block copolymers offer versatility in the structures they can form owing to the wide variety of different monomers, the different ways that the monomers can interact, and the control of the lengths of the monomer chains\cite{Bates1999}.
Several recent papers use this versatility to choose polymers in such a way as to form quasicrystal approximants, for example by choosing different monomer types in diblock micelles\cite{Gillard2016,Schulze2017}, in giant surfactants\cite{Yue2016}, in polymer liquid crystal systems\cite{Zeng2023}, and in $ABC$ triblock copolymers\cite{Chang2020}. 
The potential diversity of resultant mesoscale structures in block copolymer systems is explained in detail in the review by Huang \textit{et al.}\cite{Huang2018}.
These structures offer the potential of developing materials with unusual photonic bandgap behavior\cite{Vitiello2014, Sinelnik2020}

There is experimental evidence that quasicrystals are associated with the presence of two length scales in the system\cite{Hayashida2007,Huang2018} in a wide range of examples going beyond soft matter and materials science (for example, fluid dynamics\cite{Edwards1994,Kudrolli1998,Arbell2002} and nonlinear optics\cite{Pampaloni1995}). 
The presence of different length scales is qualitatively clear in the structure of the components for several of the soft-matter systems that form QCs, including micelles with a soft corona\cite{Zeng2004,Smart2008} and star copolymers with arms of different lengths\cite{Hayashida2007,Chernyy2018}.
The connection between having two length scales and the stability of QCs is supported by a large body of theoretical work, including from the fields of fluid dynamics and pattern formation\cite{Edwards1994,Zhang1996,Lifshitz1997,Porter2004,Lifshitz2007a,Rucklidge2012,Skeldon2015,Castelino2020}, phase field crystals\cite{Achim2014,Jiang2015,Subramanian2016,Jiang2017,Savitz2018,Jiang2020,Tang2020,Liang2022}, classical density functional theory of interacting particles\cite{Barkan2011,Archer2013,Archer2015,Ratliff2019,Subramanian2021,Scacchi2020,Walters2018}, molecular dynamics\cite{Engel2007a,Barkan2014}, and self-assembly of hard particles\cite{Damasceno2012,Je2021} and hard particles with shoulder potentials\cite{Dotera2014,Dotera2017}.
At the most basic level, the theoretical work attributes the stability of QCs to the nonlinear three-wave interaction of waves of density fluctuations on the two length scales.
For example, when the ratio of those length scales is $2\cos15^{\circ}\approx1.93$, the nonlinear interactions between two waves of one length scale and one of the other favor density waves that are spaced $30^{\circ}$ apart in Fourier space\cite{Edwards1994,Kudrolli1998,Silber2000,Arbell2002,Rucklidge2009,Skeldon2015}, giving twelve-fold symmetry. 
These arguments suggest that the length scales should usually be within a factor of two of each other to encourage QCs, with a ratio of $1.93$ for twelve-fold QCs and $1.618$ for ten-fold or icosahedral QCs\cite{Bak1985,Lifshitz1997,Lifshitz2007a,Rucklidge2009,Rucklidge2012,Subramanian2016}, with other ratios stabilizing other quasicrystals\cite{Archer2022}. 
Nevertheless, stable QCs can also be found with larger ratios, for example, an eight-fold quasipattern was found in a reaction--diffusion problem with a length scale ratio of~$4$\cite{Castelino2020}.

The theoretical arguments attribute the stability of QCs to the presence to two length scales, but this presence is not sufficient for the formation of QCs: even with two length scales, hexagons, lamellae or other structures of different sizes can be stable.
Nor is the presence of two length scales necessary for the formation of QCs: in fluid dynamics, there are examples of quasipatterns in Faraday wave experiments\cite{Christiansen1995} whose stability can be explained in the context of a single length scale\cite{Zhang1996,Rucklidge2009}.
Nonetheless, the presence of two length scales in an appropriate ratio is strongly associated with the stability of QCs\cite{Savitz2018}, though some tuning of parameters is usually needed to ensure that QCs are favored over competing crystalline phases such as hexagons and lamellae\cite{Ratliff2019}.

In this paper, we focus on what features in the polymer design can lead to two length scales in the instability towards phase separation.
We will work in the weak segregation limit, and use the Random Phase Approximation (RPA) to characterize the length scales that emerge, concentrating on the point of phase separation of block copolymer melts.
This complements other theoretical approaches to this problem, for example using self-consistent field theory\cite{Li2010,Xu2013,Tyler2007}.
The advantage of using the simpler RPA theory is that it allows a rapid search through parameter space for likely candidates of two length scale phase separation.
The disadvantage is that the theory does not predict which structure will ultimately be stable.

Prior work in this area\cite{Nap2001,Nap2006,Kuchanov2006} considered a limited range of architectures giving rise to two length scales.
Here we extend their work by (i)~working within the same classes of architecture but increasing the explored parameter space, (ii)~extending the investigation to include copolymers formed by random reaction, and (iii)~considering three component star polymers of the form investigated by Hayashida \textit{et al.}\cite{Hayashida2007}.
One focus, especially within themes (i) and (ii) above, has been to find structures that are as simple as possible to synthesise whilst retaining the two length scale feature.

\begin{figure*}[t]
 \centering
 \begin{subfigure}[t]{0.6\linewidth}
 \includegraphics[width=1\textwidth]{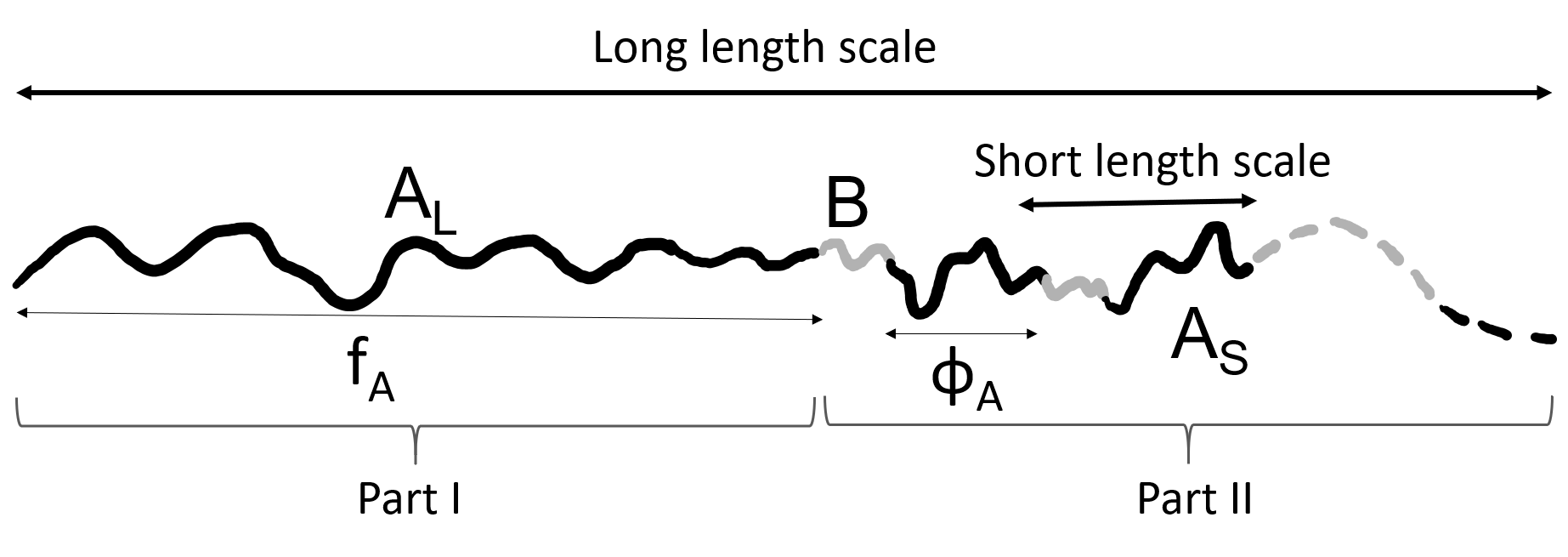}
 \caption{}
 \end{subfigure}
 \begin{subfigure}[t]{0.38\linewidth}
 \includegraphics[width=1\textwidth]{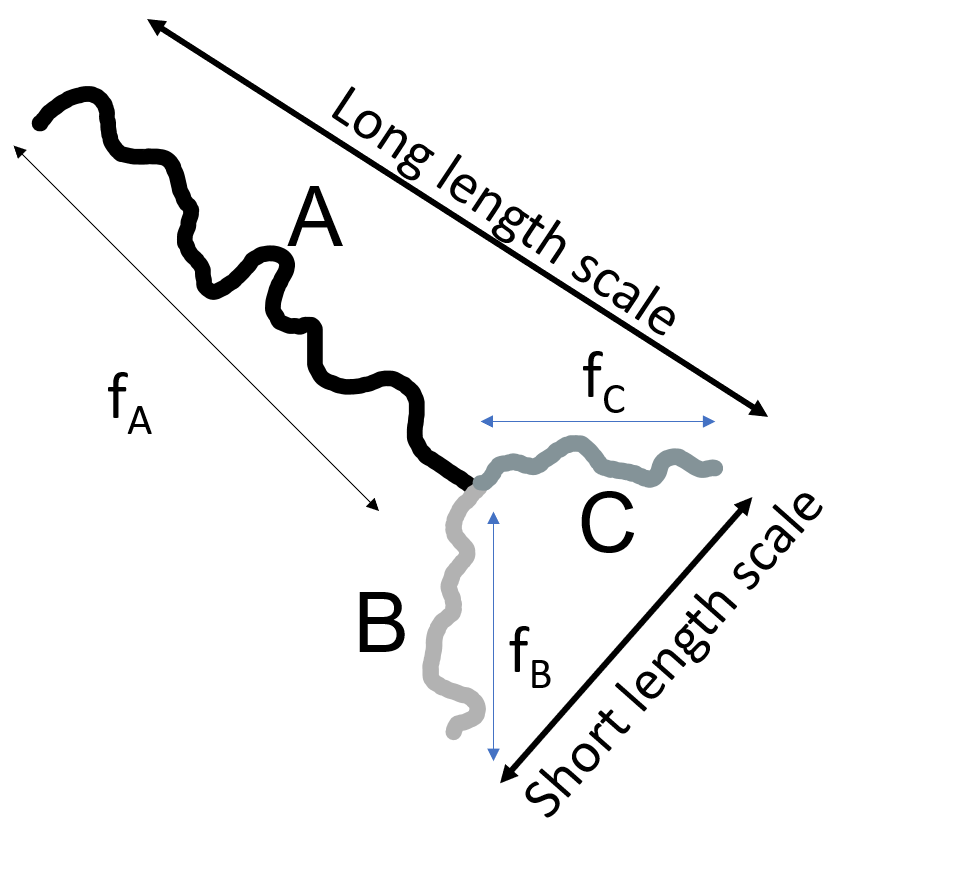}
 \caption{}
 \end{subfigure}
 \caption{Schematic models for the block copolymers.
 (a)~$A_L(BA_S)_n$ with $A$ in black and $B$ in gray, and $L$ and $S$ indicating the long and short $A$~blocks. The length fraction (of the total polymer length) of the long $A$ block in part~{I} is $f_{A}$, and within part~{II}, $\phi_A$~is the length fraction of the short $A$ blocks within each of the~$n$ $BA_S$ diblocks.
 (b)~$ABC$ star block copolymer with $A$ in black, $B$  in light gray and $C$ in mid-gray, having length fractions $f_{A}$, $f_{B}$ and $f_C$
 respectively.}
 \label{fig:models}
\end{figure*}

We consider two classes of polymer architectures, both chosen to allow two length scales to emerge, see \figurename~\ref{fig:models}, and ask whether there is one or two length scales, and in the latter case, how can the ratio of these length scales be controlled?
The first class has two types of monomer ($A$ and~$B$), while the second has three ($A$, $B$ and~$C$). 
Each example is specified by the proportions of the different components and the strengths of the interactions between them.
The selection of length scales during phase separation involves a balance between the entropy of stretching the polymer chain and the energy penalty of having incompatible monomers interacting.
Qualitatively, the phase separation lengthscales are set by the size of subsections of the chain that repel one another. To achieve phase separation simultaneously at two lengthscales requires fine tuning of the relative degrees of repulsion via composition and interaction parameters.  Typically, one requires greater average repulsion (per monomer) to drive phase separation at the short lengthscale, because the stretching energy is larger; conversely, less repulsion is needed at the longer lengthscale.  These qualitative features are present in all examples explored below.

Linear block copolymers are much easier to manufacture compared to branched block copolymers, so our first class of polymer (\figurename~\ref{fig:models}a) explores the design of linear block copolymers.
Block copolymers that are manufactured will normally exhibit polydispersity, but we start with a discussion of idealized monodisperse models before introducing some aspects of polydispersity via random assembly of the blocks.
A similar architecture was considered by Nap\cite{Nap2001}, but they restricted their investigation to cases where the $A_S$ and $B$ blocks are of equal length; here we show that relaxing that constraint leads to greater flexibility in the design space and the possibility of easier synthesis.
The monodisperse chains have a long section~$A_L$ of $A$-type monomers followed by $n$~alternating $BA_S$ diblocks, with shorter stretches of $B$-type and $A$-type monomers linked back to back. 
Microphase separation of the long $A_L$ block and the $(BA_S)_n$ tail gives one length scale, while the incompatibility between $A$ and $B$ sub-blocks within the tail can lead to microphase separation on a second length scale.
The presence of $A$ monomers in the $BA_S$~tail reduces the incompatibility between it and the $A_L$ section.
Polydispersity is achieved by starting with a mixture of $A_L$, $A_S$ and $B$ blocks.
These are allowed to react in such a way that each $B$ block links to an $A$ block at either end, the $A_S$ blocks link to $B$ blocks at either end, while the $A_L$ blocks only react with $B$ blocks at one end.
The result is a mixture of polymers of different lengths, starting and ending with $A_L$ blocks but with different lengths of $BA_S\dots{A_SB}$ blocks in between, consistent with the proportions in the initial mixture.

The second class of polymers (\figurename~\ref{fig:models}b) is an $ABC$-star structure, with different lengths of the $A$, $B$ and $C$ arms, inspired by the polymers used by Hayashida \textit{et al.}\cite{Hayashida2007}
Here, two length scales can emerge if one arm ($A$) is longer than the other two ($B$ and~$C$), with microphase separation between $A$ and $B$ with $C$ together leading to the long length scale, and microphase separation between $B$ and $C$ leading to the short length scale.
Again, this class has been explored in the literature\cite{Kim1993}; the new perspective here is the focus on phase separation with two length scales.

In each case, we explore the parameter ranges in which two length scales emerge, and we indicate the parameters for which the ratio between the two length scales would favor twelve-fold quasicrystals.
This work is part of a long-term effort to develop design criteria for polymers that will robustly and spontaneously form quasicrystals. 

Our main tool is the Random Phase Approximation (RPA) for polymer blends, which is the truncation of the free energy functional at the quadratic term in density fluctuations\cite{deGennes1970,Leibler1980}. 
The theory describes the point at which there is a transition from a polymer melt to a phase separated structure, near the point of the initial segregation. 
As a result, the theory does not identify the final stable phase.
The method uses coarse-graining of monomers into monomer units with effective bond length (or Kuhn length)~$b$, which allows the polymer to be considered as consisting of flexible and freely rotating units that can be described as a random walk.
Microphase separation occurs when there are density fluctuations that decrease the free energy of the homogeneous melt, and the wavenumber of these fluctuations gives the preferred length scale of the resulting phase separated structure.
The melt phase is (meta-)stable when a (small) density fluctuation of any wavenumber increases the free energy.

In the case of an incompressible melt with two monomer types $A$ and~$B$, and in the absence of any specific interaction between the monomer types other than incompressibility interactions, the RPA is concerned with the non-interacting structure factor~${S_{0}(q)}$ of the homogeneous and isotropic melt.
This non-interacting incompressible structure factor depends on wavenumber~$q$ and is expressed in terms of correlations between small random composition fluctuations, as described in more detail below.

Interactions between the monomer units are parameterised by the Flory interaction parameter~$\chi_{AB}$, which is related to the interaction energies between the different monomer types.
Including these interactions leads to additional composition correlations that are described in terms of the structure factor~${S(q)}$.
This structure factor also gives the form of the expected results of scattering in experiments that detect composition fluctuations in the melt.

In an incompressible two-component ($A$ and $B$) copolymer, consisting of lengths of $A$ units joined to lengths of $B$ units, possibly with repetition or branching, and with a Flory interaction parameter~$\chi_{AB}$, the structure factor~$S$ is related to the non-interacting structure factor~$S_0$ by\cite{Rubinstein2003}
 \begin{equation}
 \label{eq:basicABstructurefactor}
 S(q)=\left(\frac{1}{S_{0}(q)}-\frac{2\chi_{AB}}{\Omega\rho}\right)^{-1},
 \end{equation}
where $\Omega$ is the system volume and $\rho$ is the monomer unit density, so $\Omega\rho$ is the total number of monomer units.
When there are no interactions ($\chi_{AB}=0$), the two structure factors are the same.
The non-interacting structure factor~$S_{0}(q)$ is positive and may have a maximum at a particular wavenumber, so $S_{0}(q)^{-1}$ may have a positive minimum.
As interactions are introduced (as $\chi_{AB}$ increases), $S(q)$~goes to infinity at the value of $\chi_{AB}$ for which the term in brackets in eq~\eqref{eq:basicABstructurefactor} first goes to zero, and so phase separation occurs on the length scale corresponding to the wavenumber at which $S_{0}(q)$ is maximum.
If $S_0(q)$ has two maxima of equal height at different wavenumbers, eq~\eqref{eq:basicABstructurefactor} indicates that phase separation will happen simultaneously at both wavenumbers.

There are more general expression for compressible systems and for copolymers with more than two types of monomer. In these cases, the RPA is formulated by writing the free energy as a quadratic functional of the composition fluctuations, with multiple Flory interaction parameters.
For the melt to be stable, this quadratic form needs to be positive definite, so phase separation occurs when eigenvalues of the quadratic form change sign.

Read\cite{Read1998} formulated a method to find the noninteracting structure factor for an arbitrary block copolymer melt, using `self-terms', `coterms' and `propagator terms' to describe the architecture of the block copolymer, tracing the arrangement and connections between blocks in the chain.
Here, the self-terms describe the density correlations between monomer units within a single block, and the coterms describe the density correlations between monomers units from two different blocks on the same polymer.
The chain between two different blocks may have other blocks in between, depending on the architecture of the polymer, and the propagator terms specify how the correlations (given by coterms) are modified by the presence of the intermediate blocks.
We use this method to compute the structure factor for our two classes of copolymer architecture.

\section{Two component linear chain with fixed architecture} 

In our two component block copolymer system shown in \figurename~\ref{fig:models}(a), we index each polymer chain within the system by~$\alpha$, so $1\leq\alpha\leq n_c$, where $n_c$~is the number of chains.
Within each chain~$\alpha$, we index the location of each monomer unit by~$l$, with $1\leq l\leq N$, where $N$~is the number of monomer units in each chain, assumed (in this section) to be the same for each chain. 
The location of the monomer unit is thus~$\ralpha{r}$, and each monomer unit is of type~$A$ or type~$B$.
The density of $A$~monomer units in physical space is then a sum of delta functions, summed over all chains and all locations on each chain of the $A$ monomer units, and similarly for~$B$.
The Fourier transforms of these two density distributions are then
 \begin{equation}
 \begin{aligned}
 \label{eq:rho_Fourier_transforms}
 \rho_{\bm{q}}^{A}&=\sum_{A}\exp(i\bm{q}\cdot\ralpha{r}),\\
 \rho_{\bm{q}}^{B}&=\sum_{B}\exp(i\bm{q}\cdot\ralpha{r}),  \end{aligned}
 \end{equation}
where $\bm{q}$ is the wavevector, and the sums are taken over the locations of the $A$ and $B$ monomer units.
The structure factors for the $A$ and $B$ monomer units in the absence of any interactions are given by the correlations in the densities of the two types, which in Fourier space can be written as
 \begin{equation}
 \begin{aligned}
 \label{eq:methdeq34}
 S_{0}^{AA}(q)&=\langle\rhoqA{-\bm{q}}{A}\rhoqA{\bm{q}}{A}\rangle_0, \\
 S_{0}^{AB}(q)&=\langle\rhoqA{-\bm{q}}{A}\rhoqA{\bm{q}}{B}\rangle_0=S_{0}^{BA}(q),\\
 S_{0}^{BB}(q)&=\langle\rhoqA{-\bm{q}}{B}\rhoqA{\bm{q}}{B}\rangle_0,
 \end{aligned}
 \end{equation} 
where the angle brackets with subscript~$0$ represent the average over all possible configurations of the polymers in the absence of any interactions between monomers, within the constraints of the polymer architecture.

The incompressibility constraint requires $\rhoqA{\bm{q}}{A}=-\rhoqA{\bm{q}}{B}=\rho_{\bm{q}}$, so we need only use $\rho_{\bm{q}}$ in place of~$\rhoqA{\bm{q}}{A}$ and $-\rho_{\bm{q}}$ in place of~$\rhoqA{\bm{q}}{B}$.
Using the RPA\cite{Leibler1980}, we can express the incompressible structure factor in
the absence of interactions as:
 \begin{equation}
 \label{eq:S0qABn}
 S_{0}(q)=\frac{S_{0}^{AA}S_{0}^{BB}-(S_{0}^{AB})^2}{S_{0}^{AA}+S_{0}^{BB}+2S_{0}^{AB}}
 \end{equation}
and the free energy functional up to second order in density fluctuations (in units of $k_B T$) as: 
 \begin{equation}
 \label{eq:FrhoqABn}
 F\lbrace\rho_{\bm{q}}\rbrace =
 \frac{1}{2}\sum_{\bm{q}}\rho_{\bm{q}}\rho_{-\bm{q}}
 \left(\frac{1}{S_0(q)} - \frac{2\chi_{AB}}{\Omega\rho}\right).
 \end{equation}
If the term in brackets in eq~\eqref{eq:FrhoqABn} is positive for all wavenumbers, the melt is (meta-)stable, while if this term changes sign, the melt is unstable.
So, as in the discussion of eq~\eqref{eq:basicABstructurefactor}, the length scale of phase separation is associated with a maximum of $S_0(q)$. 

In eq~\eqref{eq:FrhoqABn}, $\Omega\rho$ is equal to the total number of monomer units, and with $n_c$ chains each having $N$ monomer units, we have $\Omega\rho=n_cN$.
We see below that $S_0(q)$ is proportional to $n_cN^2$, so by treating the Flory interaction parameter~$\chi_{AB}$ in combination with~$N$, all dependence on~$N$ can be isolated into~$N\chi_{AB}$.

 \begin{figure}[t]
                \begin{centering}
                        \includegraphics[scale=1]{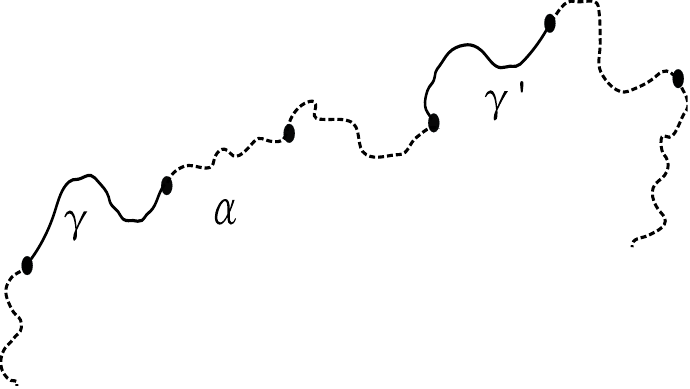}
                        \caption{We index each polymer chain within the system by~$\alpha$, and $\gamma$ and $\gamma'$ are two blocks within the chain (these could be $A_L$, $A_S$ or~$B$).}
                        \label{fig:arbitrary_chain}
                \end{centering}
        \end{figure}

We use the method of Read\cite{Read1998} to calculate the term~${S_{0}^{AA}(q)}$, ${S_{0}^{AB}(q)}$ and~${S_{0}^{BB}(q)}$ in the absence of any interaction between the monomers, for the $A_L(BA_S)_n$ structure,
previously considered by Nap\cite{Nap2001}, and shown in \figurename~\ref{fig:models}(a).
The method treats each polymer chain~$\alpha$ by splitting it into different blocks, with each block being of a single monomer type.
We illustrate this in \figurename~\ref{fig:arbitrary_chain}, with~$\gamma$ and~$\gamma'$ indicating two different blocks.
Each block is associated with a `self-term'~$J_{\gamma}$, a `coterm'~$H_{\gamma}$ and a `propagator term'~$G_{\gamma}$\cite{Read1998}.
The self-term for an individual block~$\gamma$ gives the contribution to the structure factor from that block, and comes from summing over monomer pairs within that block.
The contribution from the interaction between two blocks~$\gamma$ and~$\gamma'$ is the coterm for block~$\gamma$ multiplied by the propagator terms of all the blocks on the unique connecting path between $\gamma$ and~$\gamma'$ (see \figurename~\ref{fig:arbitrary_chain}), and finally multiplied by the coterm for block~$\gamma'$ at the end. 
This contains all pairwise monomer interactions between monomers in the two blocks $\gamma$ and~$\gamma'$.
The structure factor for a given polymer chain~$\alpha$ is then a combination of the sum of the self-terms for each block and a double sum of the product of coterms with appropriate propagators over all non-identical pairs of blocks in that chain.
 
Each block~$\gamma$ has its own normalised wavenumber~$Q_{\gamma}$, dependent on the number of monomer units~$N_\gamma$ in the block:
 \begin{equation}\label{eq:normalised_wavenumber}
        Q_{\gamma}^2 =\frac{N_{\gamma}b^2}{6}q^2,
 \end{equation} 
where $q=|\bm{q}|$.
This wavenumber is scaled by the effective bond length~$b$; for the sake of simplicity we take same length~$b$ for all blocks.
Each of the self-, co- and propagator terms are functions of these normalized wavenumbers~$Q_{\gamma}$ for each block~$\gamma$.
These terms are all based on Debye functions\cite{deGennes1970} and are given by\cite{Read1998}
\begin{align}
 \begin{aligned} \label{eq:Read_terms}
    \text{self-term:}\quad&
        J_{\gamma}=N_{\gamma}^{2}j(Q_\gamma^2)
        \hspace{1cm} \text{  where }
        j(Q^2)=\frac{2}{Q^{4}}(\exp{(-Q^{2})}-1+Q^{2}),\\
    \text{coterm:}\quad&
        H_{\gamma}=N_{\gamma} h(Q_{\gamma}^2)
        \hspace{1cm} \text{  where }
        h(Q^2)=\frac{1}{Q^{2}}(1-\exp{(-Q^{2})}),\\
        \text{propagator term:}\quad&
        G_{\gamma}=\exp(-Q_{\gamma}^{2}).
 \end{aligned}
\end{align} 

Turning now to the specific $A_{L}(BA_{S})_n$ architecture,
the polymer chain is considered in two parts.
Part~{I} is the $A_{L}$ block with length fraction $f_A$, and part~{II} is the tail $(BA_S)_n$, comprising $n$ $BA_S$~diblocks, with length fraction $1-f_A$.
Within each $BA_S$ diblock, the $A_S$ block has a length fraction of $\phi_A$ and hence length fraction of the $B$ block is $1-\phi_A$.
If the total number of monomer units in a chain is~$N$ then the number of $A$ and $B$ units in each block of the polymer chain can be written down:
\begin{equation}
        \begin{aligned}
        \label{2compeq0}
        \text{Number of $A$ monomer units in the } A_L \text{ block in part~{I}}: \quad&
        N_{A_L}= f_A N; \\
        \text{Number of $A$ monomer units in each } A_S \text{ block in part~{II}}: \quad &
        N_{A_S} = \frac{1}{n} (1-f_A) \phi_A N; \\
        \text{Number of $B$ monomer units in each } B \text{ block in part~{II}}: \quad &
        N_B = \frac{1}{n} (1-f_A) (1-\phi_A) N.
        \end{aligned}
        \end{equation}
We note that the two types of $A$ chain have the same lengths when $N_{A_L}=N_{A_S}$, or
\begin{equation}
\label{eq:ALequalsAS}
\phi_A = n\frac{f_A}{1-f_A}.
\end{equation}
For the specific $A_L(BA_S)_n$ polymer, the three normalised wavenumbers~$Q_{A_{L}}$, $Q_{A_{S}}$ and $Q_{B}$ are given by
\begin{equation}\label{eq:2comp_norm_wavenumbers}
        \begin{aligned}
        Q_{A_{L}}^{2} &=f_AQ^2,\\
        Q_{A_{S}}^{2} &=\frac{1}{n}(1-f_A)\phi_AQ^2,\\
        Q_{B}^{2}     &=\frac{1}{n}(1-f_A)(1-\phi_A)Q^2,
        \end{aligned}
        \end{equation} 
where $Q^2=\frac{Nb^2}{6}q^2$.

To calculate the structure factor, we start by treating parts~{I} and~{II} separately, and we illustrate the calculation for the case $n=2$ before discussing the case of general~$n$.

In part~{I}, there is only the $A_LA_L$ self-term:
        \begin{equation}
        {J_{AA}^{I}} 
             = {J_{A_L}} 
             = N_{A_L}^2 \, j(Q_{A_L}^2)
             = N^2 f_A^2 \, j(f_A Q^2).
        \end{equation}
There are no $J_{BB}^{I}$ and $J_{AB}^{I}$ self-terms. 
We have given the explicit dependence on $N_{A_L}$ and $Q_{A_L}$, and on $N$ and $Q$, in this case, but we will suppress this below. 
We note that, here and below, all terms and the final expressions can written as functions of the scaled wavenumber~$Q$, and that the self-terms, and the coterm--propagator term--coterm combinations, will be proportional to~$N^2$.

In part~{II}, we work out composite self-terms for the $(BA_S)_2=BA_SBA_S$ chain: $J_{AA}^{II}$, $J_{BB}^{II}$ and $J_{AB}^{II}$.
For $J_{AA}^{II}$, each $A_S$ block can interact with itself, yielding a self-term $\jA$ multiplied by~$2$ since there are two of these.
Each $A_S$ block can also interact with the other $A_S$ block: for this $A_SA_S$ interaction, we use a coterm ($\hA$ at each end) and a propagator term ($\gB$) to jump across the $B$~block.
There is a factor of~$2$ since the interaction since either $A_S$ block could be the starting point. 
Putting these together results in
        \begin{equation}
        J_{AA}^{II} = 2\jA + 2\hA\gB\hA.
        \end{equation}
Similarly, $J_{BB}^{II}$ is calculated by considering the two self-terms $\jB$ and the coterm--propagator term--coterm chain in each direction:
        \begin{equation}
        J_{BB}^{II} = 2\jB + 2\hB\gA\hB.
        \end{equation}
The last term within part~{II} is~$J_{AB}^{II}$, starting with an $A_S$ block and ending with a~$B$ block.
In this case, we have coterms $\hA\hB$ for every instance of adjacent $A_S$ and $B$~blocks, and we have coterm--propagator term--propagator term--coterm chains for the $A_S$ and $B$~blocks separated by the other $A_S$ and $B$~blocks:
        \begin{equation}
        \jAB = 3\hA\hB + \hB\gA\gB\hA.
        \end{equation}
There is an equal expression for $\jBA$, starting with a $B$ block and ending with an~$A_S$ block.

Finally, we consider interactions between parts~{I} and~{II}. The~$A_L$ block has the self-term $J_{A_L}$ and coterm $H_{A_L}$, and will interact with the $A_S$~blocks and $B$~blocks in part~{II}.
The $AA$ interactions lead to contributions $2\hAL\gB\hA$ and $2\hAL\gB\gA\gB\hA$, with the factor $2$ in each case since the $AA$ interactions can start with $A_L$ or~$A_S$. 
The $AB$ interactions, starting with $A_L$ and ending with~$B$, are $\hAL\hB$ and $\hAL\gB\gA\hB$.

These contributions are combined to give the three terms:
\begin{equation}     \label{eq:S_bits_n_2}
\begin{aligned}
    S_{0}^{AA} &= n_c\left(J_{AA}^{I} + 
              J_{AA}^{{II}} +
              2\hAL\gB\hA + 2\hAL\gB\gA\gB\hA\right),\\
    S_{0}^{AB} &= n_c\left(J_{AB}^{{II}} +
              \hAL\hB + \hAL\gB\gA\hB\right),\\
    S_{0}^{BB} &= n_c \jBB,
\end{aligned}
\end{equation}
where we have multiplied by the number of chains~$n_c$.

In the case of general~$n$, the part~{I} term remains the same.
For the other terms, we end up with longer expressions involving additional powers of $\gA\gB$ propagator terms.
After some combinatorics, and summing the resulting finite geometric series, the general~$n$ composite self-terms for part~{II} are
        \begin{equation}\label{2compeq3a}
        \begin{aligned}
        \jAA &= n\jA +
                2\hA^{2}\gB\left(
                  \frac{n\left(1-\gA\gB\right)-\left(1-(\gA\gB)^{n}\right)}
                       {\left(1-\gA\gB\right)^{2}}
                       \right),\\
        \jBB &= n\jB + 2\hB^{2}\gA\left(
                  \frac{n\left(1-\gA\gB\right)-\left(1-(\gA\gB)^{n}\right)}
                       {\left(1-\gA\gB\right)^{2}}
                           \right),\\
        \jAB &= \hA\hB\left(
                \frac{(2n+1)\left(1-\gA\gB\right)-2+(\gA\gB)^{n}+(\gA\gB)^{n+1}}
                     {\left(1-\gA\gB\right)^{2}}
                      \right).
        \end{aligned}
        \end{equation}
There is an equal expression for $\jBA=\jAB$. Then, the three terms in the total structure factor, for $n_c$ chains, are:
        \begin{equation}
        \begin{aligned}
        \label{eq:2compeq6}
        S_{0}^{AA}&=n_c\left(
                     J_{AA}^{{I}} + 
                     \jAA + 
                     2\hAL\hA\gB\left(
                          \frac{1-(\gA\gB)^{n}}
                               {1-\gA\gB}\right)
                               \right),\\
        S_{0}^{BB}&=n_c\jBB,\\
        S_{0}^{AB}&=n_c\left(
                     \jAB + 
                     H_{A}^{{I}}\hB\left(
                         \frac{1-(\gA\gB)^{n}}
                              {1-\gA\gB}\right)
                       \right).
        \end{aligned}
        \end{equation}
These three terms are proportional to $n_cN^2$ and are functions of the scaled wavenumber~$Q$. The three terms are then combined to give the overall noninteracting incompressible struture factor $S_{0}(q)$ using eq~\eqref{eq:S0qABn}, also proportional to $n_cN^2$ and a function of~$Q$.

 \begin{figure}[t]
 \includegraphics[scale=1]{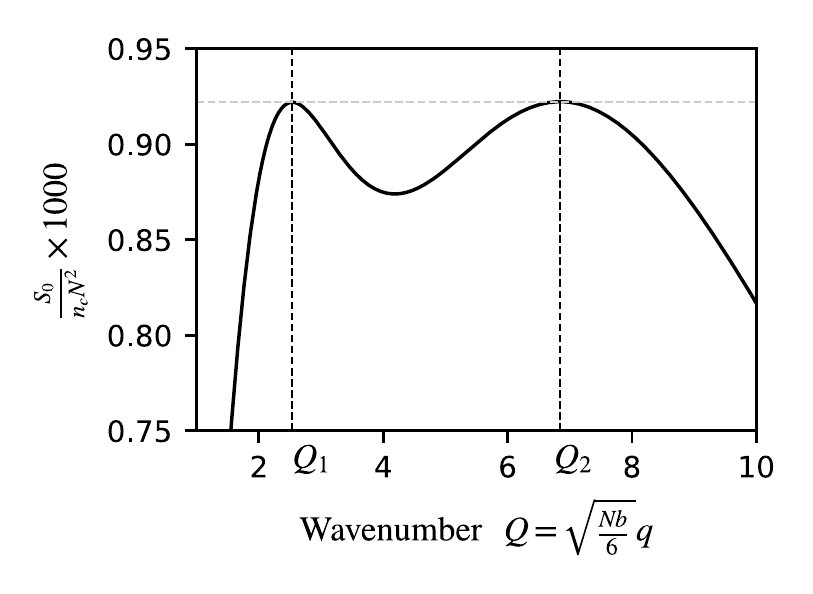}
 \caption{The incompressible structure factor $S_{0}$ (scaled by $n_cN^2$) from eq~\eqref{eq:S0qABn} as a function of scaled wavenumber~$Q$ for $f_A=0.39225$, $\phi_{A}=0.85$ and~$n=5$.
 The two maxima in the plot indicate the two length scales of phase separation.
 In this example, the wavenumbers at the maxima are $Q_1=2.53$ and $Q_2=6.85$, which have a ratio of $2.70$, and the maxima are at the same height.}
 \label{fig:scattering}
 \end{figure}

Typically, for phase separation on one length scale, the structure factor plot will have a single dominant peak.
We have chosen the architecture $A_L(BA_S)_n$ in order to allow phase separation with two length scales, which will manifest as two maxima, at scaled wavenumbers $Q_1$ and~$Q_2$ in the structure factor plot.
An example of the resulting structure factor $S_{0}$ is plotted in \figurename~\ref{fig:scattering} with $f_A=0.39225$, $\phi_{A}=0.85$ and~$n=5$.
In this example, the structure factor has two maxima at the same height, and the wavenumbers at these two maxima are in the ratio~$2.70$.
The two maxima will in general have different heights, with the higher one indicating the wavenumber that will appear first in phase separation. 
We define the ratio of wavenumbers $Q_{r}=Q_2/Q_1$, with $Q_1<Q_2$. 

\begin{figure}[t]
 \centering
 \includegraphics[width=0.9\textwidth]{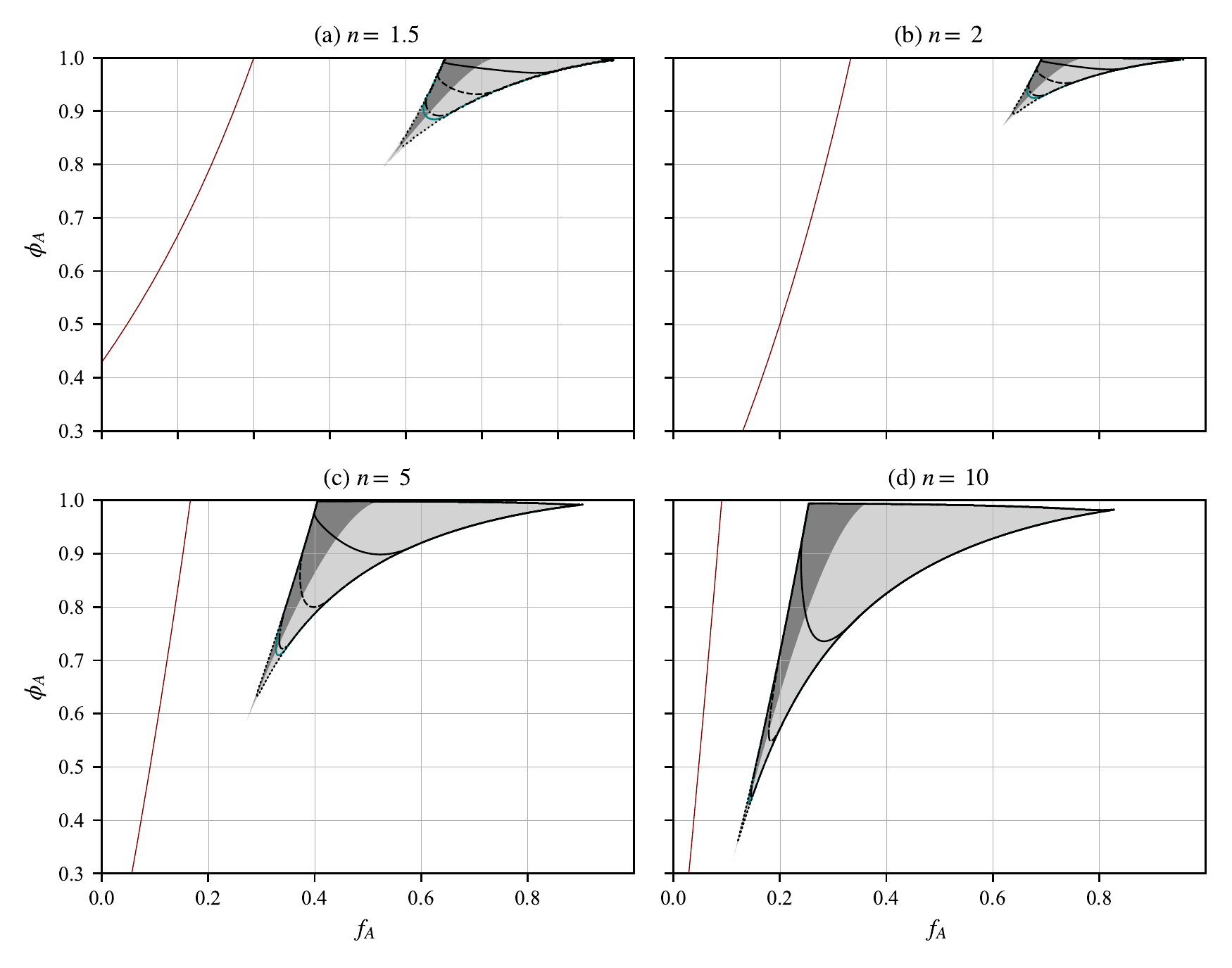}
 \vspace{-3ex}
 
 \caption{Regions of a single maximum (white) and two maxima (shaded), as a function of  $f_A$ and $\phi_{A}$ with $n=1.5$, $2$, $5$ and~$10$.
 In the shaded regions, the darker (resp.\ lighter) areas are where the maximum with the smaller (resp.\ larger) wavenumber is higher.
The solid contour line across the cusp indicates a wavenumber ratio $Q_{r}=3.5$, with dashed, dash-dot and dotted lines indicating wavenumber ratios of $2.5$, $2.0$ and $1.5$ respectively.
The teal line indicates parameter where the wavenumber ratio is~$1.93$. The maroon line is where the two types of $A$~blocks have the same length, using eq~\eqref{eq:ALequalsAS}.}
 \label{fig:contour}
\end{figure}

Whether there is one maximum or two maxima, and their heights and values of the wavenumbers at the maxima, depends on the model parameters $f_A$, $\phi_{A}$ and~$n$.
We have computed the structure factor for all possible combinations of $f_A$ and $\phi_{A}$, with $1\leq n\leq10$.
A summary of the results is presented in \figurename~\ref{fig:contour}.
As happen in other systems with transitions between one and two length scales, the boundary between the regions in parameter space separating one from two length scales are cusp-shaped\cite{Nap2001,Kuchanov2006,Bentley2021}.
Within the cusps (shaded), there are two maxima in the non-interacting structure factor and hence two length scales in the phase separation.

The architecture with $n=1$, $A_LBA_S$, does not give any two length scale phase separation. 
The smallest model in the $A_L(BA_S)_n$ family that gives a cusp is that with $n=2$ (\figurename~\ref{fig:contour}b), which is a linear chain with only $5$~blocks.
From \figurename~\ref{fig:contour}(b,c,d), we see that the area of the cusps increases with~$n$, and the maximum ratio between the two length scales increases as well.
This is because the short length scale is set by the size of the $BA_S$ blocks, while the long length scale is set by the overall size of the polymer, which increases with~$n$. 
The smallest linear chain with two components that gives two length scale phase separation is $A_LBA_SB$, indicated by $n=1.5$ in \figurename~\ref{fig:contour}(a), with the structure factor computed along the same lines as described above.

We also show in \figurename~\ref{fig:contour} the lines given by eq~\eqref{eq:ALequalsAS}, where the two $A$~blocks have the same lengths.   
We note that these lines do not intersect the cusps, which shows that having $A$ blocks of different lengths is needed to have two length scales at the point of phase separation.

\section{Two component linear chain with random assembly}

\begin{figure}[t]
\includegraphics{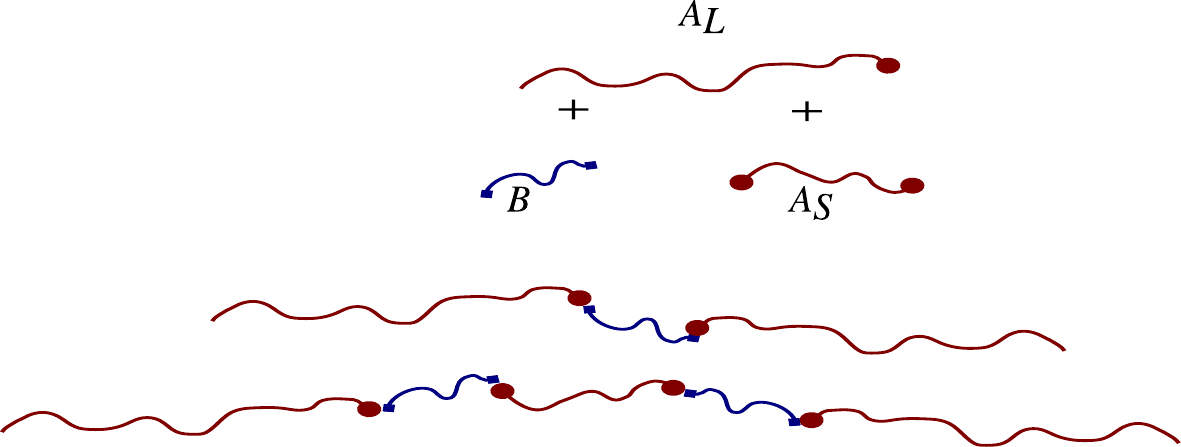}
 \caption{Schematic representation of the polydisperse model.
 Starting with a mixture of $A_L$, $A_S$ and $B$ (top), the final mixture 
(bottom) will contain polymer chains of architecture $A_{L}BA_{L}$, $A_{L}BA_SBA_{L}$, etc.}
\label{fig:ALBLexamples}
\end{figure}

The polydisperse model involves the random assembly of $A_L$, $A_S$ and $B$ blocks using the Markov chain method proposed by Read\cite{Read1998}.
In this model, the $A_L$~blocks have one reactive end, the $A_S$~blocks have two reactive ends, and the $B$~blocks also have two reactive ends. 
The $A$ reactive ends combine with $B$ reactive ends to form linear chains.
At the end of polycondensation, assuming a complete reaction and stoichiometry, the mixture will contain only chains that have $A_L$ blocks at both ends and different lengths of $BA_S\dots{A_SB}$ blocks in between, for example, $A_LBA_L$, $A_LBA_SBA_L$, etc., as illustrated in \figurename~\ref{fig:ALBLexamples}.
 
The polycondensation process starts with initial block fractions $\beta_{A_L}$, $\beta_{A_S}$ and $\beta_B$ for $A_L$, $A_S$ and $B$ blocks respectively, with $\beta_{A_L}+\beta_{A_S}+\beta_{B}=1$.
In the reaction, all the $B$ type ends will react with $A$ type ends, so the initial mixture must contain the same number of each type. 
Thus 
\begin{equation}\label{eq:polycondensation_linear_relation}
2 n_{blocks}\beta_B=n_{blocks}\beta_{A_L}+ 2n_{blocks}\beta_{A_S},
\end{equation}
where $n_{blocks}$ is the total number of blocks.
Eliminating~$\beta_B$, the condition for complete reaction is
\begin{equation}\label{block fraction condition}
        3\beta_{A_L}+4\beta_{A_S}=2.
\end{equation}
In order to use the RPA for this system, we need the number of monomer units in each block.
We use the number of monomer units in the $A_L$ block ($N_{A_L}$) to scale the numbers in the $A_S$ and $B$ blocks ($N_{A_S}$ and $N_{B}$), using scaling factors $\nu_{A_S}$ and $\nu_{B}$:
\begin{equation}\label{monomer units scaling}
        \begin{split}
                N_{A_S}&=\nu_{A_S}N_{A_L},\\
                N_{B}&=\nu_{B}N_{A_L}.
        \end{split}
\end{equation}
Thus the three parameters that describe the model are $\beta_{A_L}$, $\nu_{A_S}$ and~$\nu_{A_S}$.

Polymerization of such a mixture is by random assembly.
At each stage in the polymerization, an $A$ reactive end is always followed by a $B$ block, while a $B$ reactive end combines with $A$ reactive ends from $A_L$ or $A_S$ blocks with probability given by the proportion of the two types.
In terms of the model parameters, the probability that a $B$~block will be followed by an $A_L$~block is $P_{A_LB}=\frac{\beta_{A_L}}{2\beta_{B}}$, and that a $B$~block will be followed by an $A_S$~block is $P_{A_SB}=\frac{\beta_{A_S}}{\beta_B}$, so from eq~\eqref{eq:polycondensation_linear_relation}, we have $P_{A_LB}+P_{A_SB}=1$.
The probabilities that a $B$ block follows $A_L$ or $A_S$ blocks are both~$1$, and the probabilities of other combinations (for example, $A_S$ followed by $A_L$) are zero.

This is a Markov process, where the block that gets attached depends only on the last block, and the structure factor can be computed by the methodology of Read\cite{Read1998}.
The ideas are an extension of those discussed above, but with infinite rather than finite geometric series.
For example, for the contribution to the $S_{0}^{AA}$ part of the non-interacting structure factor from the $A_L$ to $A_L$ interactions, there are two coterm $H_{A_L}$ factors for each end, as well as sums of propagator terms corresponding to all the possible chains that can go between the two ends, weighted by the probability of that chain.
So, a single $B$ block contributes $P_{A_LB}G_B$, a $BA_SB$ chain contributes $P_{A_SB}G_BG_{A_S}P_{A_LB}G_B$, a $BA_SBA_SB$ chain contributes $(P_{A_SB}G_BG_{A_S})^2P_{A_LB}G_B$, and so on.
The infinite geometric series, with common factor $P_{A_SB}G_BG_{A_S}$, can readily be summed, for this and for the other interactions\cite{Joseph2023}. 
The outcome of these calculations is
\begin{equation}\label{in matrix form}
        \begin{split}
        S_{A_LA_L}&=N_{A_L}\Omega\rho\frac{\beta_{A_L}}{\beta_{A_L}+\nu_{A_S}\beta_{A_S}+\nu_{B}\beta_{B}} \left(j_{A_L}+h_{A_L}^2\frac{P_{A_LB}G_{B}}{1-P_{A_S}BG_{A_S}G_{B}}\right),  \\
S_{A_LA_S} &=N_{A_L}\Omega\rho\frac{\nu_{A_S}\beta_{A_L}}{\beta_{A_L}+\nu_{A_S}\beta_{A_S}+\nu_{B}\beta_{B}} h_{A_L}h_{A_S}\frac{P_{A_SB}G_B}{1-P_{A_SB}G_{A_S}G_{B}},\\
S_{A_LB}&=N_{A_L}\Omega\rho\frac{\nu_{B}\beta_{A_L}}{\beta_{A_L}+\nu_{A_S}\beta_{A_S}+\nu_{B}\beta_{B}} h_{A_L}h_{B}\frac{1}{1-P_{A_SB}G_{A_S}G_{B}},\\
S_{A_SA_S}&=N_{A_L}\Omega\rho\frac{\nu_{A_S}^2\beta_{A_S}}{\beta_{A_L}+\nu_{A_S}\beta_{A_S}+\nu_{B}\beta_{B}} \left(j_{A_S}+2h_{A_S}^2\frac{P_{A_SB}G_B}{1-P_{A_SB}G_{A_S}G_{B}}\right),\\
S_{A_S B}&=N_{A_L}\Omega\rho\frac{\nu_{A_S}\nu_{B}\beta_{A_S}}{\beta_{A_L}+\nu_{A_S}\beta_{A_S}+\nu_{B}\beta_{B}} 2h_{A_S}h_{B}\frac{1}{1-P_{A_SB}G_{A_S}G_{B}},\\
S_{BB}&=N_{A_L}\Omega\rho\frac{\nu_{B}^2\beta_{B}}{\beta_{A_L}+\nu_{A_S}\beta_{A_S}+\nu_{B}\beta_{B}} \left(j_{B}+2h_{B}^2\frac{P_{A_SB}G_{A_S}}{1-P_{A_SB}G_{A_S}G_{B}}\right).
\end{split}
\end{equation}
The overall structure factors $S_{0}^{AA}$, $S_{0}^{BB}$ and $S_{0}^{AB}$ are
\begin{equation}
        \label{Structurefactors}
        \begin{split}
                S_{0}^{AA}&= S_{A_LA_L}+2S_{A_LA_S}+S_{A_SA_S}\\
                S_{0}^{BB}&=S_{BB}\\
                S_{0}^{AB}&=S_{A_L B}+S_{A_S B}
        \end{split}
\end{equation}
These are combined into the non-interacting structure factor expression in eq~\eqref{eq:S0qABn}.

\begin{figure}[t]
        \includegraphics[trim=30 20 50 35,clip,width=0.80\textwidth]{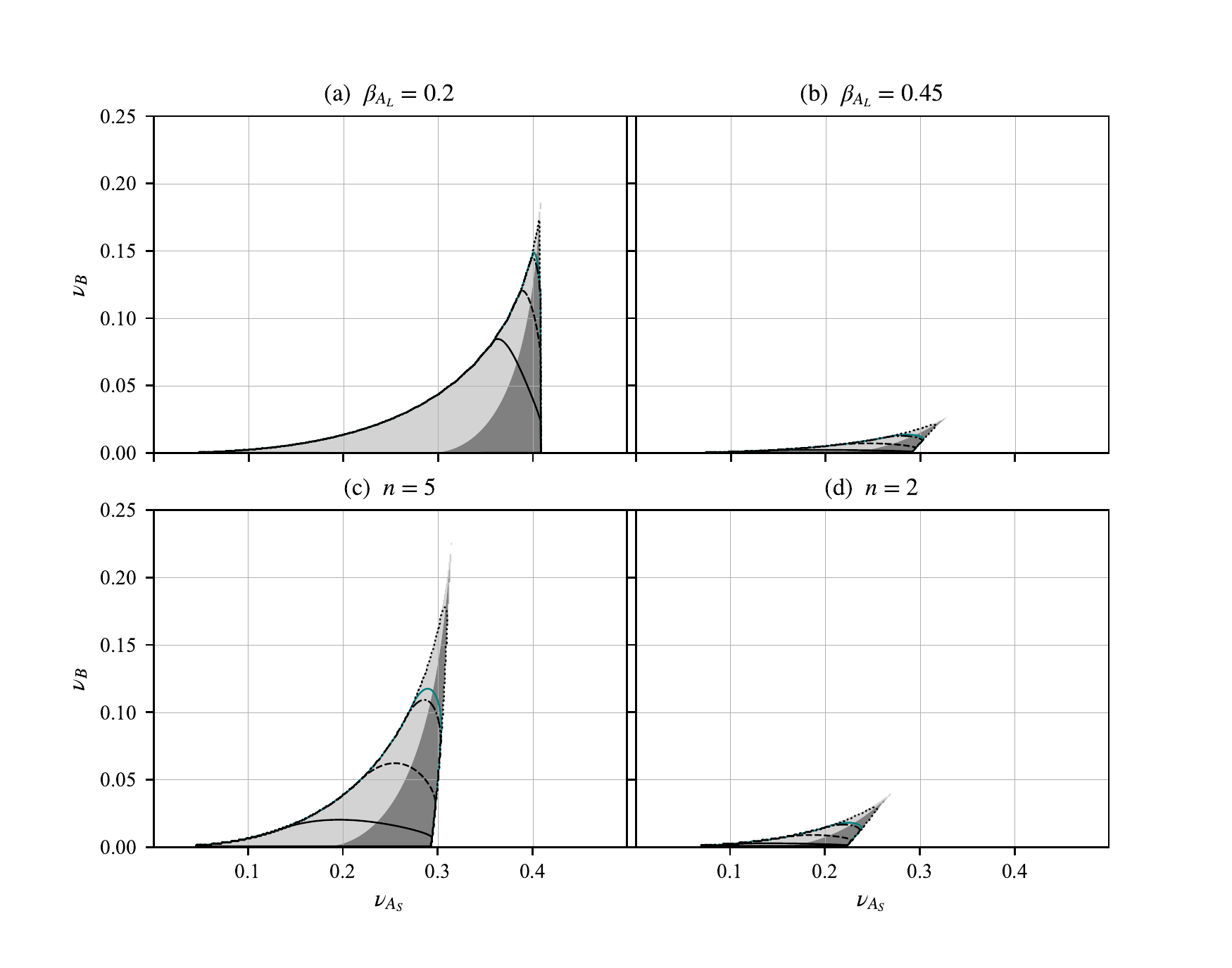}
 \caption{Top row: regions of single maxima (white) and two maxima (shaded) as a function of $\nu_{A_L}$ and $\nu_{B}$ with $\beta_{A_L}=0.2$ and $0.45$, for the two component linear chain with random assembly.  
 In the shaded regions, the darker (resp.\ lighter) areas are where the maximum with the smaller (resp.\ larger) wavenumber is higher.
The solid contour line across the cusp indicates a wavenumber ratio $Q_{r}=3.5$, with dashed, dash-dot and dotted lines indicating wavenumber ratios of $2.5$, $2.0$ and $1.5$, respectively.
The teal line indicates where the wavenumber ratio is $1.93$.
Bottom row: we plot the results for the monodisperse versions with (c)~$n=5$ and (d)~$n=2$ in terms of equivalent values of $\nu_{A_L}$ and~$\nu_{B}$. }
        \label{fig:contours_in_nu}
\end{figure}

Thus, the non-interacting structure factor $S_{0}(q)$ is determined as a function of the normalized wavenumber and the three parameters: the block fraction~$\beta_{A_L}$ and the monomer fractions~$\nu_{A_S}$ and~$\nu_{B}$. 
For fixed~$\beta_{A_L}$, the regions of $(\nu_{A_S},\nu_{B})$ where there are two maxima in the structure factor are once again cusp-shaped (see \figurename~\ref{fig:contours_in_nu} for  $\beta_{A_L}=0.2$ and $\beta_{A_L}= 0.45$).
The lines across the shaded part of the cusps indicate the ratio between the wavenumbers where the two peaks occur.
When $\beta_{A_L}$ is $0.2$ (with $\beta_{A_S}=0.35$ and $\beta_B=0.45$), there are more $A_S$ and $B$~blocks that $A_L$~block, so the polymer will form longer chains, and the region for two length scale phase separation is larger than with $\beta_{A_L}=0.45$ ($\beta_{A_S}=0.1625$ and $\beta_B=0.3875$).
In both these cases, the length scale ratio corresponding to $12$-fold symmetry is indicated by the teal line.

Direct mapping between the monodisperse fixed~$n$ and polydisperse random assembly versions of the two component linear chain models is not possible, because of the range of chain lengths possible in the second case, and because of the difference in architecture.
But, roughly speaking, $\beta_{A_L}$ is inversely proportional to the number of $A_S$ and $B$ blocks available in the polydisperse model, which is related to the number $n$ of $BA_S$ diblocks in the monodisperse case.
However, for fixed $n$ in the monodisperse case, it is possible to convert the parameters from $(f_A,\phi_A)$ in that case to $(\nu_{A_S},\nu_{B})$, but with keeping the $n$ the same.
The two length scale regions in the polydisperse case (\figurename~\ref{fig:contours_in_nu}, top row, with $\beta_{A_L}=0.2$ and $\beta_{A_L}=0.45$) are comparable to the two length scale regions in the monodisperse case (\figurename~\ref{fig:contours_in_nu}, bottom row, with $n=5$ and $n=2$).
This supports the hypothesis that longer chains and the presence of $A_S$ blocks encourage  two length scale phase separation.

\section{Three component $ABC$ star}

For a polymer system with three components and with Fourier transformed monomer densities $\rhoqA{{\bm{q}}}{A}$, $\rhoqA{{\bm{q}}}{B}$ and $ \rhoqA{{\bm{q}}}{C}$ for $A$, $B$ and $C$ monomer units respectively, the procedure for calculating the partition function is similar to the two component system, though imposing the incompressibilty condition is more involved.
The non-interacting (compressible) structure factor matrix~$M_q$ is a three-by-three matrix whose components are $S_{0}^{AA}$, \dots, $S_{0}^{CC}$:
        \begin{equation}
        \label{eq:ABC_Mq}
        M_q=\begin{bmatrix}
        S_{0}^{AA} &S_{0}^{AB} &S_{0}^{AC}\\
        S_{0}^{AB} &S_{0}^{BB} &S_{0}^{BC}\\
        S_{0}^{AC} &S_{0}^{BC} &S_{0}^{CC}
        \end{bmatrix} 
        =
        n_c N^2\begin{bmatrix}
        s_{0}^{AA} &s_{0}^{AB} &s_{0}^{AC}\\
        s_{0}^{AB} &s_{0}^{BB} &s_{0}^{BC}\\
        s_{0}^{AC} &s_{0}^{BC} &s_{0}^{CC}
        \end{bmatrix},
        \end{equation}
where, as discussed in the case of the two component system, the $S_0^{IJ}$ terms are proportional to $n_cN^2=\Omega\rho N$. With this scaling, the structure factor terms $s_0^{IJ}$ depend only on the scaled wavenumber and on parameters that describe the architecture of the polymer.

We need the inverse of this matrix, which we write as
        \begin{equation}
        \label{eq:inverse_Mq}
        M_{q}^{-1}=\frac{1}{\Omega\rho N}\begin{bmatrix}
        \Gamma^{AA} &\Gamma^{AB} &\Gamma^{AC}\\
        \Gamma^{AB} &\Gamma^{BB} &\Gamma^{BC}\\
        \Gamma^{AC} &\Gamma^{BC} &\Gamma^{CC}
        \end{bmatrix}.
        \end{equation}
Then, the free energy functional is: 
\begin{equation}
\label{eq:ABC_free_energy_3by3}
 F\left(\lbrace\rhoqA{{\bm{q}}}{A},\rhoqA{{\bm{q}}}{B},\rhoqA{{\bm{q}}}{C}\rbrace\right)
 =
 \frac{1}{2}\sum_{\bm{q}}\begin{bmatrix}
        \rhoqA{{\bm{q}}}{A} & \rhoqA{{\bm{q}}}{B} &\rhoqA{{\bm{q}}}{C} 
        \end{bmatrix}
        \left(M^{-1}_{q} +
        \frac{1}{\Omega}\begin{bmatrix}
        V_{AA}& V_{AB} & V_{AC}\\
        V_{AB}& V_{BB} & V_{BC}\\
        V_{AC}& V_{BC} & V_{CC}
        \end{bmatrix}\right)
        \begin{bmatrix}
        \rhoqA{{-\bm{q}}}{A} \\\rhoqA{{-\bm{q}}}{B} \\\rhoqA{{-\bm{q}}}{C} 
        \end{bmatrix},
\end{equation}
where the three-by-three matrix of $V_{AA}$ etc.\ expresses the interaction potential between the different monomer types, and will be written in terms of $N$ times the Flory interaction parameters below.

The incompressibility in the system provides the constraint
        \begin{equation}\label{eq:ABC_incompressibility}
        \rhoqA{{\bm{q}}}{C}= -(\rhoqA{{\bm{q}}}{A} +\rhoqA{{\bm{q}}}{B} ),
        \end{equation}
which can be used to reduce the three-by-three matrices in the free energy to two-by-two:
\begin{equation}
\label{eq:ABC_free_energy_2by2}
 F\left(\lbrace\rhoqA{{\bm{q}}}{A},\rhoqA{{\bm{q}}}{B}\rbrace\right)
 =
 \frac{1}{2}\sum_{\bm{q}}\begin{bmatrix}
        \rhoqA{{\bm{q}}}{A} & \rhoqA{{\bm{q}}}{B} 
        \end{bmatrix}
        W_q
        \begin{bmatrix}
        \rhoqA{{-\bm{q}}}{A} \\\rhoqA{{-\bm{q}}}{B} 
        \end{bmatrix}.
\end{equation}
Here, $W_q$ is a $2\times2$ matrix that contains all the non-interacting structure factor terms and the interaction potentials between different types of monomers, and is defined by
\begin{equation}\begin{aligned}
        \label{eq:ABC_Wq}
        \Omega\rho N W_q
        = 
        &\begin{bmatrix}
        \Gamma^{AA}+\Gamma^{CC}-2\Gamma^{AC} & \Gamma^{AB}-\Gamma^{BC}-\Gamma^{AC}+\Gamma^{CC}\\
        \Gamma^{AB}-\Gamma^{BC}-\Gamma^{AC}+\Gamma^{CC} &  \Gamma^{BB}+\Gamma^{CC}-2\Gamma^{BC}
        \end{bmatrix}+{} \\
        & \qquad N \begin{bmatrix}
        -2\chi_{AC} &\chi_{AB}-\chi_{BC}-\chi_{AC}\\
        \chi_{AB}-\chi_{BC}-\chi_{AC} & -2\chi_{BC}
        \end{bmatrix} .
        \end{aligned}
        \end{equation}
 Here, the Flory interaction parameters $\chi$ for the monomer pairs of $A$, $B$ and $C$  blocks are defined by
        \begin{equation}
        \begin{aligned}\label{eq:ABC_Nchi}
        \chi_{AB}&=-\frac{\rho}{2}\left(V_{AA}+V_{BB}-2V_{AB}\right),\\
        \chi_{BC}&=-\frac{\rho}{2}\left(V_{BB}+V_{CC}-2V_{BC}\right),\\
        \chi_{AC}&=-\frac{\rho}{2}\left(V_{AA}+V_{CC}-2V_{AC}\right),
        \end{aligned}
        \end{equation}
where the $\chi_{AB}$, $\chi_{BC}$ and $\chi_{AC}$ describe the interaction between pairs of monomer types.
        
For the $ABC$ star architecture, the $S_{0}^{AA}$, $S_{0}^{AB}$, \dots\ terms in the $M_q$ matrix in eq~\eqref{eq:ABC_Mq} are computed using Read's method\cite{Read1998}, though only self-terms and coterms are needed as the polymer architecture is simpler than in the two component system.
For the $ABC$ star block copolymer system with $N$ monomer units, we write the length fractions of the $A$, $B$ and $C$ arms as $f_A$, $f_B$ and $f_C$ respectively, with
        \begin{equation}
        \label{3compeq1}
        f_A + f_B + f_C=1. 
        \end{equation}
The number of monomer units in each arm is then
    \begin{equation}
        \begin{aligned}
        \label{eq:ABC_number_of_monomer_units}
        \text{Number of monomer units in the } A \text{ block}: \quad&
        N_A= f_A N; \\
        \text{Number of monomer units in the } B \text{ block}: \quad&
        N_B= f_B N; \\
        \text{Number of monomer units in the } C \text{ block}: \quad&
        N_C= f_C N = (1-f_A-f_B)N.
        \end{aligned}
    \end{equation}
The non-interacting structure factor terms corresponding to the $ABC$ star architecture are 
        \begin{eqnarray}
        \label{eq:ABC_S0AA_etc}
        \begin{aligned}
        S_{0}^{AA}&=J_{A}=N_{A}^{2}j_A,\\
        S_{0}^{BB}&=J_{B}=N_{B}^{2}j_B,\\
        S_{0}^{CC}&=J_{C}=N_{C}^{2}j_C,\\
        S_{0}^{AB}&=H_AH_B=N_{A}N_{B}h_Ah_B,\\
        S_{0}^{BC}&=H_BH_C=N_{B}N_{C}h_Bh_C,\\
        S_{0}^{AC}&=H_AH_C=N_{A}N_{C}h_Ah_C.
        \end{aligned}
        \end{eqnarray}
Here, $j_A$, $j_B$, $j_C$, $h_A$, $h_B$ and $h_C$ are the self-terms and coterms for the $A$, $B$ and~$C$ branches as in eq~\eqref{eq:Read_terms}. These are functions of the normalized wavenumbers $Q_A$, $Q_B$ and $Q_C$:
\begin{equation}\label{eq:ABC_norm_wavenumbers}
        \begin{aligned}
        Q_{A}^{2} &=f_AQ^2,\\
        Q_{B}^{2} &=f_BQ^2,\\
        Q_{C}^{2} &=f_CQ^2=(1-f_A-f_B)Q^2,
        \end{aligned}
        \end{equation} 
where $Q^2=\frac{b^2 N}{6}q^2$ as before. With this, now writing wavenumber dependence in terms of~$Q$, and taking incompressibility into account, the matrix $W_Q$ (and its eigenvalues) is proportional to $\frac{1}{\Omega\rho N}=\frac{1}{n_cN^2}$, and can otherwise be expressed a function of the scaled wavenumber~$Q$, the polymer block fractions $f_A$ and~$f_B$, and the interaction parameters $N\chi_{AB}$, $N\chi_{BC}$ and~$N\chi_{AC}$. 

 \begin{figure}[t]
 \includegraphics[scale=1]{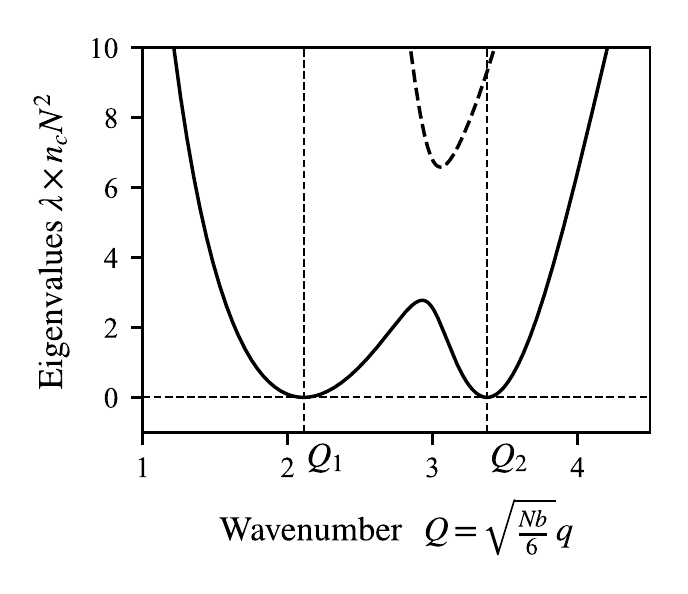}
 \caption{The two eigenvalues~$\lambda$ of $W_Q$ as functions of the wavenumber~$Q$.
 The two minima of the smaller eigenvalue (solid line) give the two length scales at phase separation.
 In this example, the wavenumbers are $Q_1=2.1096$ and $Q_2=3.3754$, with the ratio between the two wavenumbers being $Q_r=1.60$.
 The other parameters are $f_A=0.66775$, $f_B=f_C=0.166125$, $N\chi_{AB}=N\chi_{AC}=39.335$ and $N\chi_{BC}=95.070$.}
 \label{fig:eigenvalues}
 \end{figure}

With $W_Q$ defined, the quadratic form in eq~\eqref{eq:ABC_free_energy_2by2} is positive definite when the eigenvalues of~$W_Q$ are both positive, and phase separation occurs when the smallest eigenvalue of~$W_Q$ changes from positive to negative.
The wavenumber(s) at which this occurs gives the length scale(s) at phase separation.
We compute the eigenvalues of~$W_Q$ as functions of the scaled wavenumber~$Q$: see \figurename~\ref{fig:eigenvalues} for an example where the smaller eigenvalue of~$W_Q$ has two minima at zero, and so is on the verge of phase separation with two length scales in a ratio of~$1.6$. 

\begin{figure*}[t]
\begin{subfigure}[b]{0.49\textwidth}
\centering
  \includegraphics[scale=1.4]{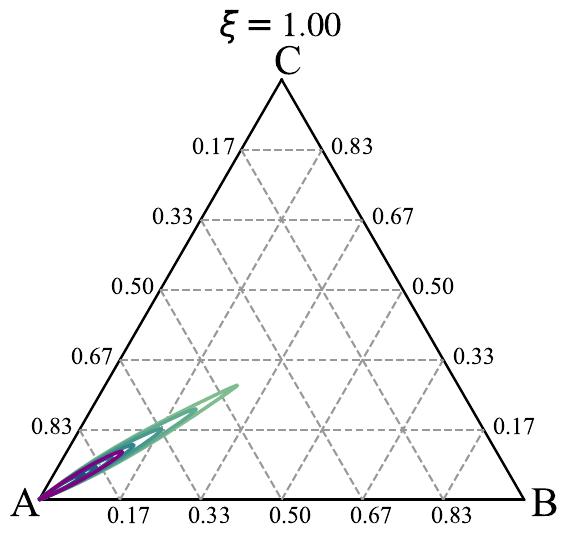}
  \vspace{2cm}
 \end{subfigure}%
\begin{subfigure}[b]{0.49\textwidth} 
 \includegraphics[scale=1]{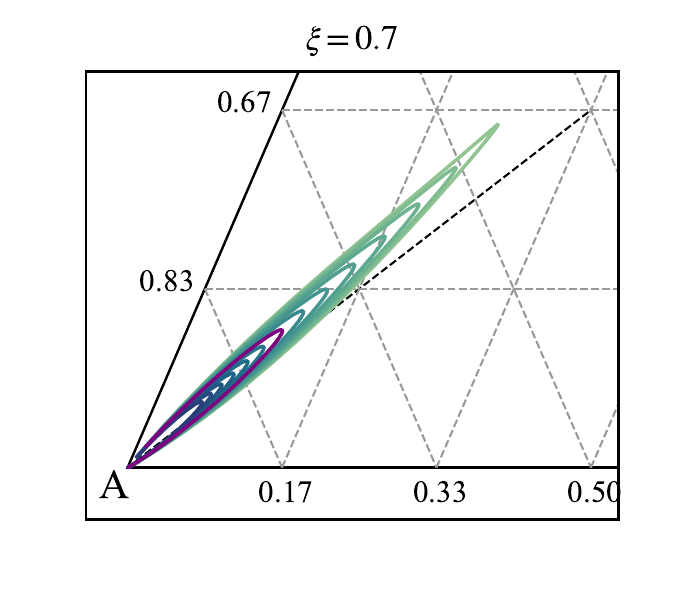}\\
 \includegraphics[scale=1]{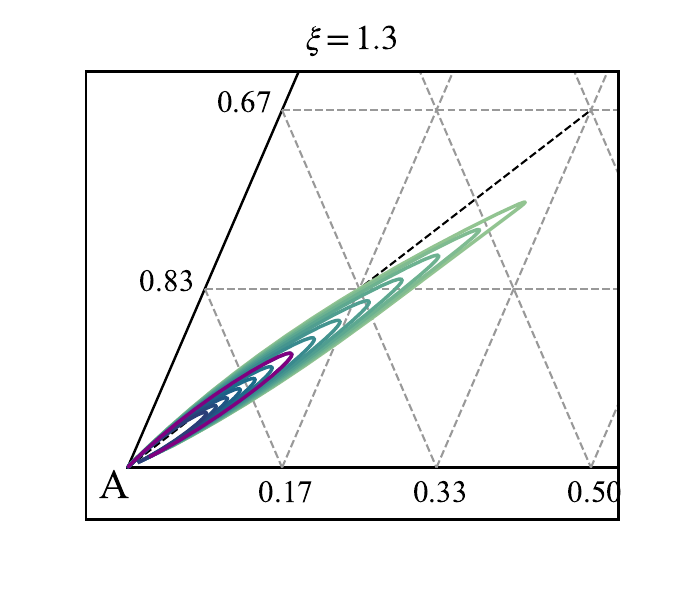}
 \end{subfigure}%
 \caption[]{Curves of constant wavenumber ratio~$Q_r$, shown as contour plots in $(f_A,f_B,f_C)$ diagrams, for three choices of~$\xi$: $\xi=1$, $0.7$ and~$1.3$.
 The outer contour in the petal shaped regions is $Q_r=1.2$, with $Q_r$ increasing to~$2.4$ in steps of~$0.2$ in the $\xi=1$ case and to~$2.5$ in steps of~$0.1$ in the other two cases.
 The purple contour is $Q_r=1.93$.}
 \label{fig:ternary}
\end{figure*}

For the three component model, the $Q$-dependent eigenvalues of~$W_Q$ depend on $5$~parameters: $f_A$, $f_B$, $N\chi_{AB}$, $N\chi_{BC}$ and~$N\chi_{AC}$, with $f_C=1-(f_A+f_B)$.  Hence, achieving instability to phase separation simultaneously at two lengthscales depends both on the ratios of the interaction parameters and on the molecular composition (this is in contrast to the two component case, where polymer composition alone determines the lengthscales of instability).
We explore this parameter space at several fixed values of the ratio of $\chi_{AC}$ to $\chi_{AB}$, defining $\frac{\chi_{AC}}{\chi_{AB}}=\xi$.
The first step is to select a desired ratio of wavenumbers~$Q_r$ (for example, $Q_r=1.6$) and a value of~$\xi$ (for example, $\xi=1$), taking $f_B=f_C$ as a starting point.  
If $\lambda(Q)$ is the smaller of the two eigenvalues, and the two minima of $\lambda(Q)$ are at wavenumbers~$Q_1$ and~$Q_2=Q_rQ_1$, the four equations for a zero double minimum of~$\lambda(Q)$, as illustrated in \figurename~\ref{fig:eigenvalues}, are:
 \begin{eqnarray}\label{eq:3compeq5a}
        \begin{aligned}
        \lambda(Q_1)=0;  \hspace{1cm} &\lambda(Q_2)=0; \\
                \frac{d\lambda}{dQ} \Bigr\rvert_{Q_1}=0;  \hspace{1cm} &        \frac{d\lambda}{dQ} \Bigr\rvert_{Q_2}=0.
        \end{aligned}
        \end{eqnarray}
We solve these four equations for~$Q_1$ and the values of the parameters $f_A$, $\chi_{AB}$ and~$\chi_{BC}$, with $\chi_{AC}=\xi\chi_{AB}$ and $f_B=f_C=\frac{1}{2}(1-f_A)$.
This allows us to identify, for the choice of $Q_r$, for the choice that $f_B=f_C$ and for the chosen ratio between $\chi_{AB}$ and $\chi_{AC}$, the other parameters at which phase separation occurs. 
We then keep the choice of $Q_r$ and the ratio between $\chi_{AB}$ and $\chi_{AC}$ fixed, but allow $f_B\neq f_C$ to compute nearby solutions to the same four equations.
In this way, we build up curves in the $(f_A, f_B, f_C)$ space on which phase separation occurs at two length scales simultaneously.
We choose other values of $Q_r$ to get several curves in that space, as illustrated in \figurename~\ref{fig:ternary}, for three choices of~$\xi$.

In \figurename~\ref{fig:ternary}, we focus on the case where the $A$ block has the longer arms, so we take $f_A>\frac{1}{3}$. 
There will be other curves of two length scale phase separation in the other corners of the triangle.
The outer contour in the petal shaped regions of two length scale phase separation is $Q_r=1.2$, with $Q_r$ increasing on the inner contours, with the larger~$Q_r$ corresponding to $f_A$ closer to~$1$.
In the case $\xi=1$, when $B$ and $C$ have the same interaction potential with~$A$, the petal region is symmetric under reflections in the $f_B=f_C$ diagonal (dotted line).
With smaller~$\xi$ ($\chi_{AC}<\chi_{AB}$) the petal bends upwards towards the $C$~region with larger~$f_C$.
Similarly, with larger~$\xi$, the petal region bends downwards towards the $B$~region.
The region of having two length scales is quite narrow, and is located in the parameter region where $A$ is long and the other two are of similar sizes.
Although we have only computed the contours up to $Q_r=2.5$, we expect that larger length scale ratios would be possible for $f_A$ closer to~$1$.

\section{Conclusion}

We have investigated the length scales that emerge at the point of phase separation in two classes of block copolymer models.
The first has alternating lengths of polymers of two different types, and the second has three different monomer types in a terpolymer star configuration.
In both cases, we find that as well as having a single length scale at phase separation, it is possible to design the polymers so that two length scales emerge.
The transition from one to two length scales occurs at a point (a cusp) in the parameter space when the length scale ratio is one.
Beyond this cusp, the length scale ratio can be made much larger than one.
In principle, the ideas and techniques developed here apply to arbitrary configurations of different types of monomers: we have explored only the simplest examples. 

We have identified experimentally accessible architectures that have phase separation at the length scale ratio~1.93 that favors twelve-fold quasicrystals. 
In the $A_L(BA_S)_n$ case, the smallest molecule has $n=1.5$, and the parameters are $(f_A,\phi_A)\approx(0.7,0.9)$ (\figurename~\ref{fig:contour}a), which corresponds to an $A_LBA_SB$ structure,
with $A_L$ being $70\%$ of the length of the chain, the two $B$'s are about $2.7\%$ each and the $A_S$ is $24.6\%$.
For molecules with more repeating $BA_S$ units, the parameters that have phase separation with two length scales have larger proportions of~$B$ and smaller proportions of~$A_L$, but these will be harder to manufacture. 
In the case of random assembly, an example choice of parameters is a mixture that is $20\%$~$A_L$, $35\%$~$A_S$ and $45\%$~$B$, where the lengths of the three chains are in a ratio $A_L:A_S:B=1:0.39:0.14$ (see \figurename~\ref{fig:contours_in_nu}a).
Of course, attention will also need to be paid to the values of $N\chi$ that are needed at the point of instability, and the relative heights of the peaks in the structure factor.
Prior work in this area\cite{Nap2001} took $\phi_A=\phi_B$; our work covers parameter values that would allow considerably easier synthesis of the polymers, in regimes that ought to favor the formation of quasicrystals.

Up until now, experimental observations in linear block copolymer melts have mainly found only relatively simple structures, such as hexagons and lamellae, or hierarchical two length scale structures, such as lamellae-within-lamellae with several layer thicknesses\cite{Faber2012,Ritzenthaler2003}.
There are indications that more complex structures can be stable: self-consistent field theory studies of linear $ABAB$ tetrablock copolymers\cite{Zhao2018} report structures including a lamella--sphere phase and a gyroid phase. 
Cylindrical 12-fold quasicrystal approximants in linear $ABCB$ terpolymer melts have also been found in self-consistent field theory calculations\cite{Duan2018}, and the Fourier transform images reported in that paper have two length scales in the characteristic 1.93 ratio.

To our knowledge, ours is the first presentation of phase separation with two length scales in the $ABC$ star terpolymer system.
This is a system where quasicrystals and close approximants have been found experimentally\cite{Hayashida2007}.
As in the $A_L(BA_S)_n$ case above, values of the three $N\chi$ parameters are quite large, suggesting that self-consistent field theory or strong segregation theory would be an appropriate next step.

Of course, our RPA calculations are only the first step: these reveal the length scales but not the final stable structures.
Subsequent steps, involving weak segregation theory\cite{Erukhimovich2005}, self-consistent field theory\cite{Li2010,Xu2013,Jiang2015b,Duan2018,Zhao2018}, strong segregation theory\cite{Gemma2002}, etc., have been carried out by several authors, but finding quasicrystals has been challenging, partly because the calculations in all these cases are quite demanding.
The final structure will be influenced by the length scale ratio and the heights of the peaks in the structure factor or values of the eigenvalues in the dispersion relation.
The examples in \figurename{}s~\ref{fig:scattering} and~\ref{fig:eigenvalues} are where these are equal at the two length scales, but the shaded regions in \figurename{}s~\ref{fig:contour} and \ref{fig:contours_in_nu} indicate where one or other peak height is larger. 
In some (phase field crystal) models of soft matter quasicrystals, fully developed quasicrystals are found where the peaks have the same or similar heights\cite{Subramanian2016,Savitz2018}, while in other (density functional theory) models, quasicrystals are found where the peaks have quite different heights\cite{Archer2013,Archer2015}.
In the models, the stability of fully developed quasicrystals depends not only on the eigenvalues at the two length scales but also on the (negative) eigenvalues at other length scales that might feature in regular crystalline orderings that compete with quasicrystals\cite{Ratliff2019}.
Our work provides this information and so should provide useful starting parameter values with two length scale phase separation, which should be a good place to start a search for quasicrystals or their approximants, experimentally or using more sophisticated theoretical methods.

\begin{acknowledgement}

The authors thank Profs Andrew Archer, Ron Lifshitz, 
An-Chang Shi and Bart Vorselaars for
stimulating conversations, and Joanna Tumelty for advice on numerical methods. 
MJ~is grateful for a PhD studentship from the Soft
Matter and Functional Interfaces (SOFI) Centre for Doctoral Training (EP/L015536/1) and the
School of Mathematics, University of Leeds. AMR~is grateful for support from
the Engineering and Physical Sciences Research Council (EP/P015611/1) and the
Leverhulme Trust (RF-2018-449/9).

The data associated with this paper are openly available from the University of Leeds Data Repository (\url{http://doi.org/10.5518/1333})\cite{Joseph2023b}.

For the purpose of open access, the authors have applied a Creative Commons Attribution (CC BY) licence to any Author Accepted Manuscript version arising from this submission.

\end{acknowledgement}



\providecommand{\latin}[1]{#1}
\makeatletter
\providecommand{\doi}
  {\begingroup\let\do\@makeother\dospecials
  \catcode`\{=1 \catcode`\}=2 \doi@aux}
\providecommand{\doi@aux}[1]{\endgroup\texttt{#1}}
\makeatother
\providecommand*\mcitethebibliography{\thebibliography}
\csname @ifundefined\endcsname{endmcitethebibliography}
  {\let\endmcitethebibliography\endthebibliography}{}

\end{document}